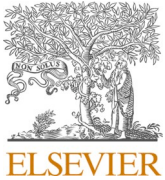

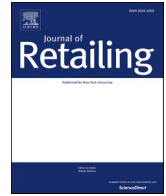

# The impact of brand equity on vertical integration in franchise systems

Mohammad Kayed [a,*], Manish Kacker [b], Ruhai Wu [b], Farhad Sadeh [c]

[a] Williams School of Business, Bishop's University, 2600 College Street, Sherbrooke, Quebec J1M 1Z7, Canada
[b] DeGroote School of Business, McMaster University, 1280 Main St. W., Hamilton, ON L8S 4M4, Canada
[c] Eastern Illinois University, 600 Lincoln Avenue Charleston, IL 61920-3099, United States

ARTICLE INFO

ABSTRACT

Brand equity and vertical integration are focal, strategic elements of a franchise system that can profoundly influence franchise performance. Despite the recognized importance of these two strategic levers and the longstanding research interest in the topic, our understanding of the interplay between brand equity and vertical integration (company ownership of outlets) in a franchise system remains incomplete. In this study, we revisit the five-decade-old question of how brand equity affects vertical integration in a franchise system and present some novel, nuanced insights into the topic. Evidence from a Bayesian Panel Vector Autoregressive model on a large panel data set shows that brand equity has a powerful, lagging inverse effect on vertical integration, such that higher brand equity leads to less downstream vertical integration in a franchise system. Reverse causality analyses identify a less pronounced but present reciprocal effect. Boundary conditions analyses reveal that the negative effect of brand equity on vertical integration is weaker in franchise systems with international presence and in retail-focused (vs. service-focused) franchises, and stronger in franchise systems with more financial resources. These findings (a) challenge traditional views (e.g., transaction cost theory, resource-based view, ownership redirection hypothesis) on the topic by demonstrating a negative effect for brand equity on vertical integration in franchise systems and showing that greater financial resources amplify this effect, and (b) shed new light on the intricate dynamics (temporal causation, reverse causation) and contingencies of this debated effect. Managerially, this research draws attention to the underrecognized strategic benefit of brand equity in mitigating channel governance issues and advise against unnecessary vertical integration, especially when brand equity is robust.

## 1. Introduction

Franchising is one of the most widespread forms of retailing. In 2022, there were more than 790,000 business establishments in the U.S. franchise systems, employing more than 8.4 million people, and with direct economic output close to \$826 billion (IFA, 2023). In pursuit of rapid growth, market expansion, operational efficiencies, and risk mitigation, many firms rely on franchising as their organizational form of choice (Dant et al., 2011; Sawant et al., 2021). One of the key determinants of the effectiveness of a firm's decision to adopt franchising as its organizational form of retailing is the way the franchise system itself is organized, i.e., the proportion of the outlets in the franchise system that is company-owned vs. franchised or the degree of downstream vertical integration[1] in the franchise system (Hsu et al., 2017; Srinivasan, 2006).

* Corresponding author.
E-mail addresses: mkayed@ubishops.ca (M. Kayed), mkacker@mcmaster.ca (M. Kacker), wuruhai@mcmaster.ca (R. Wu), fsadeh@eiu.edu (F. Sadeh).

Extant franchising research not only demonstrates how the degree of vertical integration in a franchise system impacts its overall financial performance[2] (Butt et al., 2018; Hsu et al., 2017; Srinivasan, 2006), but also emphasizes the importance of pursuing an optimal degree of vertical integration. In this regard, recent evidence demonstrates that investors react positively to franchisors' initiatives to achieve an optimal degree of vertical integration, irrespective of whether these initiatives entail a decrease or increase in vertical integration (Sadovnikova et al., 2023), underscoring the importance of operating at an appropriate degree of vertical integration for a franchise system.

Against this backdrop, an important managerial question emerges: How does one of the most influential and distinctive elements in the franchise system, i.e., brand equity (Wu, 1999), influence a franchisor's decision when it comes to vertical integration? In essence, brand equity is the cornerstone of the franchising business model and franchising can be thought of as a "leasing of the brand name" as Brickley and Dark (1987, p.402) refer to it. Furthermore, brand equity and vertical integration are two strategic levers in the franchise setup that (a) compete for constrained firm resources and (b) both have profound impact on the performance of the franchise system (Srinivasan, 2006; Hsu et al., 2017). Therefore, a sound understanding of the causal, long-term interplay between one of the most influential elements of a franchise system (brand equity) and one of the most strategically and financially consequential decisions in the franchise setup (vertical integration) is imperative to informed strategic decision making by franchise managers. Furthermore, given the inherent variability in organizational characteristics and environmental contexts across franchise systems, a proper understanding of the moderating factors that influence the relationship between brand equity and vertical integration is crucial for holistic and context-aware managerial decision making. Hence, two important managerial questions arise: (1) Does stronger brand equity lead to a higher or lower degree of vertical integration within a franchise system? And (2) What franchise- and industry-level factors influence the link between brand equity and vertical integration and how?

The importance of the abovementioned managerial questions is further accentuated by the resurging interest in vertical integration amongst franchising practitioners in particular (Kelso, 2023) and amongst retailing practitioners in general (Harper, 2020). This is surprising considering that (a) companies' track record of vertical integration had been "ugly" (Rumelt, 1974); (b) recent business memory is still replete with unsuccessful vertical integration adventures by prominent brand names (Bruno, 2021; Schmalz, 2018); and (c) financial markets have developed a habit of harshly punishing vertical integration ventures from their commencement (Moeller et al., 2005). Therefore, this marks a timely juncture to revisit the question of whether franchise systems with higher brand equity should pursue more or less vertical integration.

On the theoretical front, the question of whether stronger brand equity leads to lower or higher degree of vertical integration within a franchise system is a five-decade-old debate in the franchising literature.[3] As evident in Tables 1 and 2, previous research has used different theoretical lenses (such as Agency Theory, Transaction Cost Theory, and Signaling Theory) for examining this question, and arrived at conflicting findings (sometimes from the same theory and sometimes by the same authors). Therefore, a proper understanding of the strategic interplay between brand equity and vertical integration is crucial to the advancement of franchising theory. The theoretical significance of this cannot be further emphasized given that, as per Transaction Cost Theory (TCT), the growth of the franchising business model emerged as a pragmatic response to costly and often ineffective vertical integration (Williamson, 1991). This is why, in the TCT literature, franchising is often referred to as a "hybrid" form of channel governance that combines governance aspects of "markets" and "hierarchies" to allow firms to reap the benefits of vertical integration *without* the need for actual vertical integration (Williamson, 2005).

To inform strategic decision making by franchise managers and to help in settling the long-standing theoretic debate on the impact of brand equity on vertical integration amongst franchising researchers, we (a) review existing research on the impact of brand equity on vertical integration in franchising, (b) present a hypothesis that reflects the majority of extant empirical evidence on the topic, (c) draw on the theory of self-enforcing contracts from new institutional economics (Klein, 1985; 2002) and the power-dependence theory from social psychology (Emerson, 1962) and present a rival hypothesis, and (d) test these two hypotheses and examine boundary conditions using a Bayesian Panel Vector Autoregressive model on a large panel data set of 9151 observations from North American franchise chains.

Our results reveal that brand equity has a powerful, lagging inverse effect on vertical integration in franchise systems, such that higher brand equity leads to less downstream vertical integration. The impulse response functions (IRFs) from our empirical analyses indicate that a shock to one of the proxies of brand equity negatively impacts the degree of vertical integration in a franchise system.

---

[1] Wherever "vertical integration" is mentioned in this article, it indicates the degree of downstream or forward vertical integration, as reflected by the "proportion of company-owned outlets in the franchise system." In a similar vein, prior research in franchising has used the term "dual distribution" to refer to the "proportion of franchised outlets in the franchise system." Dual distribution has also been used to indicate the concurrence/ multiplicity of channels, such as the use of both reps and house accounts, or the use of both dealership channel and a rental agency channel. Tapered integration has also been used in the literature to indicate the presence of both owned and franchised outlets in the system.

[2] The literature shows mixed effects for vertical integration (proportion of company-owned outlets) on performance measures. For example, Kosova and Lafontaine (2010) report mixed results for the effect of vertical integration on chain survival and chain growth; Hsu et al. (2017) show that the proportion of franchised outlets has a negative impact on stock returns and idiosyncratic risk; Madanoglu and Castrogiovanni (2018) report a U-shaped effect for franchising proportion on chain survival; Srinivasan (2006) observes that dual distribution (proportion of franchised outlets) increases intangible firm value for some firms but decreases it for others depending on firm characteristics.

[3] This debate traces back to Oxenfeldt and Kelly's (1968) JR article "Will successful franchise systems ultimately become wholly-owned chains?"

**Table 1**
Empirical research on the impact of brand equity on vertical integration in franchise systems.

| Study | Proxies of Brand Equity | Method | | | | | Theory | Findings |
|---|---|---|---|---|---|---|---|---|
| | | Controlled for Endogeneity | Modeled Lagged Effects | Controlled for Unobserved Heterogeneity | Examined Causality | Investigated Reverse Causality | Theoretical Lens | Impact of BE on VI |
| Brickley and Dark (1987) | Repeat Customers | | | | | | Agency Theory | + |
| Norton (1988a) | Travel Intensity | | | | | | Transaction Cost Theory | + |
| Minkler and Park (1994) | Market Value minus Book Value | | | X | | | Transaction Cost Theory | + |
| Nickerson and Silverman (2003) | Advertising | | | | | | Transaction Cost Theory | + |
| Lafontaine and Shaw (2005) | Advertising | | | X | | | Agency Theory | + |
| Norton (1988b) | Travel Intensity | X | | | | | Agency Theory | − |
| This Study | Advertising, Brand Ranking, Franchise Fee | X | X | X | X | X | Theory of Self-enforcing Contracts, Power-Dependence Theory | − |

BE: Brand Equity; VI: Downstream vertical Integration.

**Table 2**
Theoretical predictions about the impact of brand equity on vertical integration in franchise systems.

| Theory Predicting a Positive Impact | Theory Predicting a Negative Impact |
|---|---|
| **Transaction Cost Theory**: The brand equity of a franchise system is a valuable intangible asset that must be safeguarded against franchisees' opportunism. Therefore, as brand equity increases, vertical integration should increase (Minkler & Park, 1994; Nickerson & Silverman, 2003; Norton, 1988a). | **Signaling Theory**: when brand equity is low, the franchisor owns a portion of the franchise system to signal its commitment to the brand to current and potential franchisees. Then, as brand equity increases, vertical integration decreases (Gallini & Lutz, 1992). |
| **Agency Theory**: The brand equity of a franchise system is a valuable intangible asset that must be safeguarded against franchisees' free riding. Therefore, as brand equity increases, vertical integration should increase (Brickley & Dark, 1987; Lafontaine & Shaw, 2005; Mathewson & Winter, 1985). | **Agency Theory**: when brand equity is high, a franchisor forfeits more economic gains to shirking managers than to independent franchisees. Hence, as brand equity increases, vertical integration should decrease (Norton, 1988b). |
| **Ownership Redirection Hypothesis**: Franchisors use franchisees to open markets, grow the chain, develop consumer acceptance of the brand, and then appropriate that brand equity by terminating or otherwise ending the franchisees' franchising rights. Therefore, as brand equity increases, vertical integration should increase (Oxenfeldt & Kelly, 1968; Dant et al., 1996). | **Resource Scarcity Theory**: Building brand equity and vertically integrating the channel are two strategic activities that both require major resources and impact performance. The franchisor has to allocate its constrained resources to both activities appropriately. Hence, the availability of resources for one activity (brand building) naturally limits their availability for the other (vertical integration). Therefore, as brand equity increases, vertical integration should decrease. (Gillis et al., 2014; Carney & Gedajlovic, 1991). |

The shock (a sudden increase) takes a year or two to start materially affecting the degree of vertical integration in the franchise system; however, its effect keeps building momentum over time rather than fading away. One of the notable findings from our study is the detection of bidirectional, temporal causal flows between brand equity and vertical integration. That said, the results clearly indicate that the effect of brand equity on vertical integration is more pronounced, powerful, and persistent than the reverse effect. Our results remained consistent following a battery of robustness checks and alternative model specifications. Empirical analyses of boundary conditions reveal that: (a) the negative effect of brand equity on vertical integration is weaker in franchise systems with international presence and in retail-focused franchises (in comparison with service-focused franchises), and (b) the effect is stronger in franchise systems with more financial resources, which provides another refutation of one of the most debated arguments in the franchising literature, the "ownership redirection hypothesis" (Oxenfeldt & Kelly 1968; Dant et al., 1996).

Theoretically, our study contributes to four marketing research streams. First, we contribute to the franchising and retailing literature by offering the first comprehensive examination of the causal, temporal, reciprocal, and moderated effects of brand equity on vertical integration in franchise systems. Second, we advance the broader distribution channels literature by providing empirical evidence (from the franchising context) that supports existing theoretical views on the impact of brand equity on vertical integration in distribution channels. Third, we advance the brand equity literature by highlighting the significant role brand equity plays in the governance of franchise systems, where it functions as a "safeguarding asset" that curtails franchisee opportunism, thereby reducing the need for extensive vertical integration. Fourth, we add to the marketing interactions literature (which examines interactions among marketing strategy elements) by exploring the dynamic interaction between brand and distribution strategy.

Managerially, despite the resurgence in interest in vertical integration amongst franchising and retailing practitioners, our findings advise against unnecessary vertical integration especially in situations where the franchise system enjoys a moderate to high level of brand equity. We show that as brand equity increases, franchisors lean more on their brands to safeguard themselves against franchisees' opportunism and to effectively govern their franchise systems, without the need for extensive involvement in direct distribution. Additionally, we put in the hands of brand managers of franchise systems useful empirical evidence that can help them in selling brand building initiatives to senior management and other organizational stakeholders. This research sheds light on an under-recognized governance-related benefit for brand equity in franchise systems: brand equity functions as a safeguarding asset that effectively mitigates many channel governance problems through contractual self-enforcement, which in turn alleviates the need for costly and often ineffective vertical integration. Therefore, brand investments can be seen as "dual investments" directly in the brand and indirectly in the channel – this makes their risk/reward ratio superior to many other alternatives, especially investments in direct distribution. This can make the challenging task of gaining organizational support for brand building activities much easier for brand managers of franchise systems.

## 2. Theory & hypotheses

The question of how brand equity affects vertical integration in a franchise system has attracted a reasonable amount of theoretical and empirical attention from researchers, primarily in organizational economics. Much of existing research approaches the question from a pure economic organization point of view, relying on the theoretical lenses of transaction cost theory (TCT) and agency theory (AT). The central idea here is that a franchisor's brand equity is an intangible specific asset that must be safeguarded against potential franchisee opportunism and/or free-riding. Opportunistic franchisee behaviors (deviations from contractual terms or agreed standards), such as compromising on quality, failing to comply with brand standards, selective product offerings against franchise guidelines, or unauthorized use of intellectual property, pose significant threats to the brand image and reputation of the franchise. Similarly, franchisee free-riding behaviors (reliance on the assets and resources of the franchise without making the expected contributions), such as insufficient local marketing, minimal customer service, poor facility upkeep, and inadequate employee training, can erode the perceived brand value of the franchise. As brand equity increases, the risks associated with franchisee opportunism and free-riding become more pronounced and detrimental. This compels the franchisor to pursue more vertical control within the franchise system to protect brand reputation, ensure consistent brand image, and maintain perceived brand quality. This pursuit of vertical control often manifests in the form of a more hierarchical channel governance structure i.e., a higher degree of downstream vertical integration within the franchise chain. Hence, stronger brand equity calls for more vertical integration within a franchise system. In what follows, we provide a detailed discussion of this idea through the perspectives of transaction cost theory and agency theory.

Proponents of transaction cost theory perceive a franchisor's brand equity as an intangible specific asset that stimulates opportunistic behavior by franchisees. Hence, as brand equity increases, the threat of opportunism rises, and the brand owner (franchisor) rationally relies on a more hierarchical channel governance structure (higher degree of downstream vertical integration) to safeguard this valuable specific asset – the brand. For example, Norton (1988a) examines a sample of franchise chains, from the eating places and motel industries in the U.S., and finds that as brand equity increases, firms rely more on vertical integration because brand equity "creates opportunistic incentives" (Norton 1988a, p.108). In the same vein, and within the context of the U.S. trucking industry, Nickerson and Silverman (2003) observe that the more a trucking company (motor carrier) invests in its brand name, the more likely it is to employ company drivers, as opposed to owner-operators i.e., the more vertically integrated it is. Along the same lines, Minkler and Park (1994) empirically examine a sample of public American franchises from three industries (restaurants, hotels, and professional services) and find that an increase in brand equity is positively related with an increase in the degree of downstream vertical integration in the franchise system.

In a similar spirit, agency theorists view a franchisor's brand equity as a motivation for free riding by franchisees due to the inherent incentive divergence between the brand lessor (franchisor) and the brand lessee (franchisee). Hence, as brand equity increases, franchisees' incentive to free ride on the brand, by under-delivering the pledged services or lowering quality standards, increases. Therefore, an increase in brand equity, calls for more vertical control i.e., higher degree of vertical integration to mitigate the risk of franchisees' moral hazard. Brickley and Dark (1987) study a sample of American franchises and report evidence on a positive relationship between brand equity and downstream vertical integration. Similarly, Lafontaine and Shaw (2005) establish, using a multi-industry longitudinal sample of franchise chains, that companies with more valuable brand names are more vertically integrated and argue that they do so to protect their brands from franchisees' free riding. In the same vein, Mathewson and Winter (1985) demonstrate, using a game theoretic model, that when brand equity increases, franchisees' temptation to free ride on the brand name increases which consequently increases monitoring costs. In response to that, franchisors rely more on vertical integration in governing their franchise chains.

Despite the differences in the theoretical lenses and methodological approaches of the aforementioned studies, there seems to be a convergence in their conception of the subject matter. First, all these studies view the relationship between brand equity and vertical integration as a pure economic organization concern and thus approach it from a cost-centered perspective that is focused primarily on managing transaction/agency costs. Second, they perceive brand equity as a potential target or victim of opportunistic behavior and/or moral hazard and something that needs to be preserved or safeguarded against such undesirable behaviors. Therefore, with this cost-centric and protection-focused view, it is no surprise that research in this space has predicted a positive relationship between brand equity and vertical integration (see Tables 1 and 2). To represent this line of thinking we introduce the following hypothesis:

*H₁: Higher brand equity leads to more downstream vertical integration in a franchise system (higher proportion of company-owned outlets).*

In contrast with the above-discussed view (which perceives brand equity as an external transactional attribute or as "an asset to be safeguarded" against partner opportunism and/or moral hazard), and in line with a deep-rooted view in economics (Hoos, 1959; Nerlove & Arrow, 1962) and marketing (Fischer & Himme, 2017), we recognize brand equity as a dynamic, strategic asset that can be used to safeguard the firm against partners' opportunism and/or moral hazard. Then, we draw on the power-dependence theory (Emerson, 1962) and the theory of self-enforcing contracts (Klein, 1985; 2002) to establish our theoretical arguments (Okhuysen & Bonardi, 2011). Our primary argument is that the role of brand equity in a franchising relationship is too significant (Brickley & Dark, 1987; Wu, 1999) to be reduced to being a transactional attribute or a target of franchisees' moral hazard and/or opportunism. We argue that a franchisor's brand equity itself is an effective "safeguarding asset" against franchisees' opportunism and/or moral hazard, rather than "an asset to be safeguarded" against these threats, because it simultaneously and effectively functions as a reliable incentive and a credible threat when dealing with franchisees. This carrot-and-stick mechanism effectively motivates franchisees to exercise self-enforcement, which provides a strong safeguard against their opportunism and/or moral hazard and subsequently diffuses pressures for instituting vertical bureaucratic controls through vertical integration. In what follows, we provide a detailed discussion of our theoretical reasoning.

As per the power-dependence theory (Emerson, 1962), a franchisee's dependence on a franchisor is (a) directly proportional to the franchisee's motivational investment in goals mediated by the franchisor, and (b) inversely proportional to the availability of those goals to the franchisee outside the franchisor's franchise system. When a franchisee joins a franchise system, its main motivational investment is the economic benefits (e.g., enhanced cash flows, stable cash flows, access to capital, economies of scale, operational efficiencies, return on investment) and competitive advantages (e.g., market share, customer loyalty, reputational capital, supply chain synergies) provided by that franchise system (Blair & Lafontaine, 2005; Hoy et al., 2017; Hsu et al., 2017). These valuable economic and competitive benefits are not only predominantly mediated by the brand equity of the franchise system (Brickley & Dark, 1987; Hsu et al., 2017), but also limited in franchise systems without a strong brand name (Blair & Lafontaine, 2005; Wu, 1999).[4] Therefore, as the brand equity of a franchise system increases, the magnitude and certainty of these sought-after economic and competitive benefits increase and the availability of comparable benefits outside the franchise system decreases. Hence, the stronger the brand equity of a franchise system, the more dependent are the franchisees on the franchisor. This view is consistent with Davis and Mentzer's (2008) argument that brand equity increases seller's dependence on brand owners. Therefore, *the stronger the brand equity of a franchise system, the more dependent are the franchisees on the franchisor.*

As per the theory of self-enforcing contracts (Klein, 1985; 2002), performance in most business relationships is secured through contractual self-enforcement rather than legal enforcement due to inherent contractual incompleteness. In a franchising relationship, contractual self-enforcement occurs when the benefits a franchisee expects to gain from the franchising relationship are greater than those available outside. In such a case, a credible termination sanction is sufficient to motivate the franchisee to provide the desired effort level and not act opportunistically (Klein, 2002; Wathne & Heide, 2000). Thus, the dependence imbalance dynamic created by the franchisor's brand equity implies that a franchisee's long-term losses from noncompliance and the potential consequent termination of the franchising relationship significantly outweigh the short-term benefits from acting opportunistically. In addition, as the brand equity of the franchise system increases, the system's attractiveness to high-quality franchisees increases and the bargaining power of the franchisor also increases (Ghantous & Christodoulides, 2020). This makes the process of replacing a noncompliant franchisee much easier should that franchisee choose the short-term gains of opportunism over the long-term rewards of compliance; in other words, it makes the franchisor's termination sanction more credible. This credible termination sanction (the stick), coupled with the aforementioned dependence imbalance (the carrot), significantly alleviates the holdup problem for the franchisor and increases the contractual self-enforceability in the franchising relationship. Therefore, *the stronger the brand equity of a franchise system, the higher the contractual self-enforceability in the franchising relationship.*[5]

In interorganizational relationships like franchising, vertical integration (vertical bureaucratic control) is typically conceived as a solution to governance problems such as partner moral hazard, holdup, and opportunism (Klein et al., 1978; Heide, 1994; Rindfleisch & Heide, 1997). Contractual self-enforcement is considered more efficient than vertical bureaucratic control and more effective than legal enforcement in solving such governance problems (Heide & John, 1988; Klein & Leffler, 1981; Wathne & Heide, 2000). Hence, *the higher the contractual self-enforceability in the franchising relationship, the less is the need for downstream vertical integration.*

Bringing together the preceding three arguments, our theoretical view on the impact of brand equity on vertical integration can be summarized as follows: the stronger the brand equity of a franchise system, the more dependent are the franchisees on the franchisor and the more credible is the franchisor's termination sanction. This discourages franchisees from acting opportunistically and leads to a self-enforcing contractual relationship that is anchored in the rewards of compliance and the costs of deviation. This in turn alleviates the franchisor's channel governance problem and reduces the need for vertical bureaucratic control through vertical integration. To reflect our line of thinking,[6] we advance the following rival hypothesis:

$H_{1\ (alt)}$: *Higher brand equity leads to less downstream vertical integration in a franchise system (lower proportion of company-owned outlets).*

## 2.1. Boundary conditions

As evident in the aforementioned discussion, the interplay between brand equity and vertical integration within a franchise system is multifaceted, dynamic, and shaped by various considerations. Additionally, franchise systems vary in their characteristics, the challenges they face, and the environments in which they operate. This suggests that the relationship between brand equity and vertical integration is likely to be influenced by certain franchise- and industry-level factors. For instance, the challenges faced by international franchises can significantly vary from those faced by domestic franchises. Similarly, the strategic priorities of franchises with limited resources are likely to differ from those of franchises with more resources. Also, the competitive, operational, regulatory, and consumer-related challenges faced by retail-focused franchises are likely different from those faced by service-focused franchises.

---

[4] Srivastava et al. (1998) provide a detailed explanation of how brand equity translates into growing, persisting economic rents by enhancing cash flow, accelerating cash flow, reducing volatility in cash flow, and boosting the residual value of cash flow. Fischer and Himme (2017) provide a summary of existing empirical evidence on this.

[5] Kaufmann and Lafontaine (1994) found evidence that McDonald's intentionally leaves "excess rents" on the table for its downstream partners as a mechanism for countering their opportunism and incentivizing them to exercise self-enforcement. Michael and Moore (1995) report that this practice is also common among European franchisors who deliberately leave "well-above-average returns" for their franchisees as a mechanism for curbing their opportunism through self-enforcement. Furthermore, they report that these excess rents vary from one franchisor to another, with larger brands tending to leave more rents on the table for their channel partners.

[6] Existing franchising research provides arguments that support our view of the impact of brand equity on vertical integration in franchise systems (see Table 2). For example, Gallini and Lutz (1992) offer a game theoretic signaling argument suggesting that when brand equity is low, the franchisor owns a portion of the franchise system to signal its commitment to the brand to current and potential franchisees. Then, as brand equity increases, vertical integration decreases. In the same vein, Norton (1988b) argues that when brand equity is high, a franchisor forfeits more economic gains to shirking managers than to independent franchisees. Hence, as brand equity increases, vertical integration should decrease.

Such factors are expected to play a role in shaping the franchisor's vertical integration strategy. Therefore, without a boundary conditions analysis that explores the nuances of the relationship between brand equity and vertical integration, our understanding of this important relationship will remain incomplete.

Research on dual distribution in franchising highlights four key factors that can influence the relationship between brand equity and vertical integration: firm resources, international presence, chain size, and industry effects. This literature (e.g., Gallini & Lutz, 1992; Pénard et al., 2003; Srinivasan, 2006) draws attention to the franchisor's dual pursuit of chain growth (expanding through additional franchised outlets) and effective governance (maintaining an optimal degree of vertical control through company-owned outlets), underscoring the importance of examining the moderation effects of firm resources and chain size on the link between brand equity and vertical integration. Also, it highlights the unique challenges faced by international franchises, particularly the heightened information asymmetry they encounter compared to domestic franchises, and its impact on the franchise's dual distribution strategy (Pénard et al., 2003; Wilson & Shailer, 2015). Additionally, it emphasizes the significant inter-industry variations in dual distribution dynamics, highlighting how sector-specific factors such as consumer behavior, competition dynamics, regulatory requirements, and operational challenges can influence the franchisor's vertical integration strategy (Gallini & Lutz, 1992; Pénard et al., 2003; Srinivasan, 2006). In what follows, we describe the expected moderation effects of the aforementioned four factors, discuss their theoretical underpinnings, and advance our moderation hypotheses.

First, as per resource scarcity theory (see Table 2), building brand equity and vertically integrating the channel are two strategic activities that compete for the franchisor's constrained resources (Gillis et al., 2014; Carney & Gedajlovic, 1991). Therefore, the negative relationship between brand equity and vertical integration is expected to be weaker in franchise systems with more resources. The ownership redirection hypothesis (Oxenfeldt & Kelly 1968; Dant et al., 1996) further supports this view, suggesting that franchisors initially rely on franchised outlets when resources are limited. As resources become more abundant, franchisors tend to transition toward company ownership to better manage and control the brand while allocating resources more efficiently. Hence,

**H₂**: *The negative effect of brand equity on vertical integration is weaker in franchise systems with more resources.*

Second, franchising research establishes that international franchisors encounter higher environmental uncertainty and face greater heterogeneity of regulatory, competitive, and operational factors across markets than domestic franchisors (Aliouche & Schlentrich, 2011; Elango, 2007). It also suggests that a franchisor's "superior capability" to manage franchisees' opportunism under these diverse and challenging conditions is critical for global success (Shane, 1996a). In response to such complexities, international franchisors may rely more on vertical integration, as direct ownership of outlets enables closer oversight, mitigating the risks associated with depending on foreign franchisees. Furthermore, safeguarding brand equity is more critical in international contexts, where varying local interpretations and potential misalignments call for tighter operational control to ensure consistency in customer experiences and product quality (Aliouche & Schlentrich, 2011; Shane, 1996a). Finally, international franchisors face a more intricate strategic calculus, where factors such as market entry, economies of scale and scope, competition, and the unavailability of high-quality partners often outweigh other considerations, which can further reinforce the preference for direct ownership and vertical bureaucratic control (Jell-Ojobor et al., 2022). Therefore, the negative relationship between brand equity and vertical integration is expected to be weaker for international franchisors. Hence,

**H₃**: *The negative effect of brand equity on vertical integration is weaker in franchise systems with international presence.*

Third, franchising research establishes that in larger franchise chains, a lower degree of vertical integration is often associated with fewer agency problems due to reduced shirking owing to a lower proportion of company-owned outlets as well as enhanced monitoring capabilities tailored to franchisees. Shane (1996b) explains that economies of scale allow franchisors to monitor franchisees more rigorously and cost-effectively, addressing franchisee-specific agency problems such as underinvestment in system-specific assets and free-riding on system-wide efforts.[7] Furthermore, franchisors improve their ability to monitor franchisees over time through organizational learning, further reducing agency problems as the chain grows. In addition to that, franchising research provides empirical evidence of a negative relationship between chain size and vertical integration (Kacker et al., 2016; Shane et al., 2006). Thus, larger franchise chains inherently tend to operate with a lower degree of vertical integration, which can accentuate the influence of brand equity in further reducing vertical integration within these chains. Hence,

**H₄**: *The negative effect of brand equity on vertical integration is stronger in larger franchise systems.*

Finally, regarding industry effects, retail-focused franchises face some unique challenges that often compel them to exercise more vertical control over the chain compared to service-focused franchises. For instance, a critical issue in the retail sector is inventory management, where avoiding both overstocking and stockouts is crucial for the success of individual outlets, as well as for the success of the franchise system as a whole (Li et al., 2023). This might drive retail-focused franchisors to pursue more direct ownership of outlets to maintain optimal inventory control and efficient distribution. Furthermore, retail-focused franchises rely heavily on

---

[7] Unlike employees, who require monitoring to mitigate employee-specific agency problems like shirking due to fixed compensation, franchisees require distinct monitoring mechanisms to ensure alignment with the franchisor's goals. Brickley and Dark (1987) provide a comparison of franchisee- and employee-specific agency problems.

high-visibility locations to drive sales, while service-focused franchises may prioritize operational quality and customer relationships over foot traffic. As a result, the greater strategic importance of location in retail can compel retail-focused franchisors to pursue ownership of prime, high-traffic outlets more aggressively than their service-focused counterparts (Achabal, 1982; Nishida, 2017). In doing so, retail-focused franchisors can reduce dependence on franchisees owning key outlets, thereby safeguarding brand representation and ensuring operational stability. Additionally, the typically narrower profit margins in retail sectors compared to services sectors (Bhattacharya et al., 2022; Ulaga & Reinartz, 2011) can compel franchisors to retain ownership of a larger proportion of outlets when the long-term profit retention outweighs the initial capital investment. Hence,

*$H_5$: The negative effect of brand equity on vertical integration is weaker in retail-focused (vs. service-focused) franchise systems.*

## 3. Data and measurements

*Data.* The data sources we use in this study are *Bond's Franchise Guide* and the *Annual Franchise 500 Ranking* by *Entrepreneur* magazine. Both sources have been used in prior research, and their consistency and reliability have been verified by several researchers (e.g., Lafontaine, 1995; Shane et al., 2006). Researchers from various disciplines have used *Bonds'* and *Entrepreneur*'s data in their work, and some have used the two sources jointly, as we do in this study. Using these two sources, we compiled a panel data set of North American, franchise-level annual observations for the period from 2001 to 2009. Our data set is an unbalanced panel that consists of 9151 observations.

*Endogenous variables.* We measure our first endogenous variable, vertical integration, as the percentage of company-owned units in the overall franchise system. We obtain this measure by dividing the number of company-owned outlets by the total number of outlets (company-owned plus franchised) in the franchise system (e.g., Lafontaine & Shaw, 2005; Windsperger & Dant, 2006).

We recognize that our second endogenous variable, brand equity, is a complex, multidimensional construct and that, like previous research in this domain (see Tables 1 and 5), we use proxies to measure this construct. However, for a proxy to be valid, the link between the proxy and the target construct should be based on "reasonable assumptions" (Antia et al., 2017). To achieve this, we (a) rely on existing, established proxies that were used by previous research in this domain and whose link to brand equity is explicit and/or reasonable, and (b) use two proxies rather than one to reflect the multidimensional nature of the construct. Additionally, in the robustness checks section, we replace one of the proxies with an alternative third proxy.

Our first proxy is the *advertising fee,* which is an ongoing fee that is contractually imposed by the franchisor on all its franchisees for the sake of promoting the brand through advertising. This fee is in the form of a percentage of total sales that is paid periodically by each franchisee toward an advertising fund that is managed by the franchisor. Windsperger (2004, p.1364) notes that "The more important the franchisor's brand name … the more marketing investments (national advertising and promotion measures) are required to maintain the brand name value, and the higher are the advertising fees paid by the franchisees." Prior research in marketing (e.g., Agrawal & Lal, 1995; Windsperger, 2004), economics (Lafontaine & Shaw, 2005; Hussain & Windsperger, 2013), and management (Nickerson & Silverman, 2003) has employed this proxy in its operationalization of brand equity.[8]

The second proxy is *media recognition.* We measure media recognition as the reverse coded ranking of the franchise system by *Entrepreneur Magazine's Franchise 500* annual ranking of the top 500 North American franchises. *Entrepreneur* states that it uses a proprietary algorithm developed by its panel of experts to rank franchise systems based on a set of factors that include the brand. Shane and Spell (1998, p.50) maintain that when it comes to franchise systems, an "indication of brand name value is the system's ranking in Entrepreneur Magazine." In the same spirit, Combs et al. (2004) assembled a panel of experts consisting of hospitality executives and academics, asked them to rank the franchise chains in their sample, and then used this ranking as a proxy for brand equity. Rao (1994), Shane and Foo (1999) provide a detailed justification for this approach for measuring "intangible capabilities" such as brand equity. Shane and Foo (1999) provide a detailed description of the ranking process and the magazine. In the robustness checks section, we replace this proxy with "Franchise Fee" as an alternative proxy of brand equity.

*Control variables.* To account for the confounding effects of other governance mechanisms, we control for four commonly deployed franchisee governance mechanisms that are often at play within a franchise system: selection, socialization, incentives, and royalty rate (Antia et al., 2017; Butt et al., 2018). Selection and socialization are ability-influencing franchisee governance mechanisms; and incentives and royalty rate are motivation-influencing governance mechanisms (measures are illustrated in Table 3). In addition, franchising research suggests that certain firm-level factors such as firm resources, geographic scope, experience in direct distribution, and chain size can influence the structure of a franchise system (Hennart, 2010; Wang et al., 2020; Windsperger & Dant, 2006). Hence, we control for the effect of firm resources using two proxies: financing support (whether the franchisor provides financing support to its current and prospective franchisees) and chain age (Dant & Kaufmann, 2003; Hsu et al., 2017). Previous research (e.g., Lafontaine, 1992; Minkler & Park, 1994) has used these two measures as indicators of firm resources based on the arguments that (a) the more established the franchise system, the higher its capital availability, and (b) a franchisor should already have substantial resources to be able to finance its franchisees. Furthermore, a franchisor's ability to extensively engage in direct distribution might be influenced by whether it possesses or lacks the required knowledge and expertise for doing so. Some franchisors do not rely heavily on direct distribution simply because they do not have the required skill and experience to do that, regardless of any other consideration, whereas others do it simply because they can. To address this, we control for the business development time, which is the period for which the franchisor operated as a non-franchising business directly dealing with end customers before licensing its first franchisee. Also, following prior research, we control for chain size as a proxy for firm performance (Shane et al., 2006). Moreover, we control for the geographic scope of the franchise system – whether the franchise is available in international markets or not (Hsu et al., 2017). Finally, there could be some systematic characteristics or prevailing trends within an industry as a whole that influence firms' behavior in that industry when it comes to vertical integration. To account for this, we control for the industry that the franchise belongs to.

---

[8]  Advertising Intensity (i.e., advertising-to-sales ratio), the equivalent of Advertising Fee in the non-franchising literature, is a widely used proxy of brand equity in various disciplines (e.g., Sridhar et al., 2016; Chang & Rhee, 2011; Bernile et al., 2018).

**Table 3**
Variables and measures.

| Variable | Symbol | Measure |
|---|---|---|
| Vertical Integration | VI | Percentage of company-owned units in the overall chain. |
| Advertising Fee | Ad | Percentage of sales contributed by the franchisees to the brand advertising fund. |
| Media Recognition | Media | *Entrepreneur Magazine's Franchise500* annual ranking coded in reverse order (501-Rank). |
| Chain Size | lnSize | The natural logarithm of the total number of outlets in the chain (franchised + company-owned). |
| Industry | Industry | A categorical variable (dummy coded) representing the industry the company operates in as classified by *Bond's*. |
| Business Development Time | BDT | Period in years between the year of business inception and the start of franchising. |
| Geographic Scope (International Presence) | International | A dummy variable that is set to 1 if the brand has one outlet overseas and 0 otherwise. |
| Chain Age | Age | The number of years from the start of franchising till the data collection year. |
| Selection | Selection | A composite index (as per Antia et al., 2017) reflecting the franchisor's self-reported emphasis on four qualification criteria (franchisee's net worth, business experience, industry-specific experience, and education). |
| Incentives | Incentives | A dummy variable that is set to "1" if the franchisor allows the franchisee to establish additional outlets within its area and "0" otherwise. |
| Socialization | Socialization | A composite index (as per Antia et al., 2017) reflecting the number of ongoing services provided by the franchisor to the franchisees (e.g., central data processing, central purchasing, training). |
| Royalty Rate | Royalty | The royalty rate paid by the franchisee to the franchisor as a percentage of sales. |
| Financing Support | Financing | A dummy variable that is set to "0" if the franchisor provides no financing option to its franchisees and "1" otherwise. |

In Table 3, we provide a summary of the measures we use, along with their symbols as they appear in the model. The descriptive statistics and correlations are available in Web Appendix A.

## 4. Econometric modeling

As discussed earlier, the impact of brand equity on vertical integration in franchise systems has been investigated by a number of researchers, primarily in organizational economics, during the past five decades. While this body of research provides valuable insights into the impact of brand equity on vertical integration in the franchising context, it also offers opportunities for a more nuanced examination of the relationship. This study builds upon that body of research by addressing some of the methodological limitations and nuances highlighted by prior studies – in particular, lagged effects, causality, endogeneity, and reverse causality.

First, in terms of lagged effects, marketing researchers have long established that only a small portion of the total effect of brand equity appears in the short run, while the majority of the effect is often realized in the long run (Aaker & Jacobson, 1994; Mizik, 2014). Considering this (and given the strategic long-term nature of most channel and brand decisions), a more nuanced understanding of the impact of brand equity on vertical integration necessitates the modeling of lagged effects. Second, the firm's decision to increase its vertical integration or to invest in the brand are two strategic decisions that could be influenced by several firm-level financial and non-financial factors. Besides, brand strategy and channel strategy are both elements of the firm's overall marketing strategy. This strategic interplay between brand and channel decisions warrants a rigorous accounting for potential endogeneity when studying the relationship. Third, our understanding of the impact of brand equity on vertical integration would be incomplete without assessing some form of causality and also examining the potential for reverse causality.

To account for the abovementioned methodological nuances and dynamics and to facilitate a deeper understanding of the impact of brand equity on vertical integration in franchise systems, this study adopts a Bayesian Panel Vector Autoregressive model with control variables (BPVARX). This methodological approach effectively addresses these nuances and dynamics and builds upon prior research to provide a more in-depth examination of the impact of brand equity on vertical integration in franchise systems (see Table 1).

### 4.1. The empirical model

*Model motivation.* To address the methodological considerations discussed earlier, we need an econometric modeling approach that enables us to (a) investigate lagged effects while controlling for endogeneity and firm-level heterogeneity, (b) estimate the "long-term or cumulative effects of causal variables" (Borah & Tellis, 2016, p.148), (c) conduct an assessment of reverse causality, and (d) "get as close to causality as possible with nonexperimental data" (Kang et al., 2016, p.72). These modeling requirements suggest the use of a Bayesian Panel Vector Autoregressive model (e.g., Canova & Ciccarelli, 2013; Chakravarty & Grewal, 2011) with control variables (BPVARX).

Panel Vector Autoregressive (PVARX) models (e.g., Borah & Tellis, 2016; Kang et al., 2016; Hewett et al., 2016) are powerful empirical models in that they bring together the ability of panel data models to capture unobserved individual heterogeneity with the dynamism of vector autoregressive models in their ability to model lagged effects while treating variables as endogenous and allowing for feedback loops among them. Bayesian Panel Vector Autoregressive (BPVARX) models bring in an additional layer of power by addressing some of the limitations of unrestricted (traditional) VAR models[9],[10].

*Model specification.* To test for the presence of unit roots that could lead to spurious regressions (Granger & Newbold, 1974), we used two tests of panel data stationarity, a Levin et al. (2002) test - which assumes a common unit root process for all variables - and an Augmented Dickey-Fuller (ADF) test (Choi, 2001) - which assumes individual unit root processes. For the first endogenous variable, *VI*, both the Levin-Lin-Chu ($p < 0.03$) and the ADF ($p < 0.0001$) tests rejected the null hypothesis of unit root presence. So, this variable enters the BPVARX system in level. A unit root was detected in the other two endogenous variables *Ad* and *Media*. Therefore, they are represented in the BPVARX model in their first differences. Next, we conducted a Johansen procedure (Johansen, 1995) to test for the presence of cointegrated vectors among the endogenous variables. The test reported no cointegrating equations by both the trace test ($p < 0.05$) and the maximum Eigenvalue test ($p < 0.05$). For lag length selection, we used all of the five common lag length selection criteria (Akaike Information Criterion, Akaike's Final Prediction Error Criterion, Schwarz Information Criterion, Hannan-Quinn information criterion, and Sequential Modified LR Test Statistic). As shown in Web Appendix B, three-out-of-five lag length selection criteria suggest the use of five lags ($L = 5$) and the other two selection criteria suggest the use of three lags ($L = 3$). Therefore, we use five lags ($L = 5$) for the specification of our model where all our endogenous variables are represented by five lags. This lag length is sufficient to eliminate any residuals correlation from the model as was further confirmed by the results of a Ljung-Box test (Ljung & Box, 1978) where the test failed to reject the null hypothesis of no serial correlation in residuals ($Q = 11.62; p > 0.23$). Additionally, in the robustness checks section, we also run the model on three lags ($L = 3$) as suggested by the other two lag length selection criteria.

*Model construction.* To explore the causal relationship between brand equity and vertical integration in franchise systems, we develop the following BPVARX model:

$$
\begin{pmatrix} VI_{it} \\ \Delta Ad_{it} \\ \Delta Media_{it} \end{pmatrix} = \begin{pmatrix} C_{VI,0} \\ C_{Ad,0} \\ C_{Media,0} \end{pmatrix} + \sum_{l=1}^{L} \begin{pmatrix} \beta_{11}^{l} & \beta_{12}^{l} & \beta_{13}^{l} \\ \beta_{21}^{l} & \beta_{22}^{l} & \beta_{23}^{l} \\ \beta_{31}^{l} & \beta_{32}^{l} & \beta_{33}^{l} \end{pmatrix} \begin{pmatrix} VI_{i,\ t-l} \\ \Delta Ad_{i,\ t-l} \\ \Delta Media_{i,\ t-l} \end{pmatrix} + \begin{pmatrix} \gamma_{1,1} & \cdots & \gamma_{1,10} \\ \vdots & \ddots & \vdots \\ \gamma_{10,1} & \cdots & \gamma_{10,10} \end{pmatrix} \begin{pmatrix} X1_{i,t} \\ X2_{i,t} \\ \vdots \\ X10_{i,t} \end{pmatrix} + \begin{pmatrix} \varepsilon_{VI,i,t} \\ \varepsilon_{Ad,i,\ t} \\ \varepsilon_{Media.i,\ t} \end{pmatrix}
$$

where $i = 1, 2, …,$ N firms is the cross-sectional index; $t = 1,2,…,$T years is the longitudinal time index; $l = 1,2,…,$L lags is the lag index; VI, Ad, Media are the endogenous variables (see Table 3); $C_0$ is the intercepts vector; β and γ are coefficients vectors to be estimated; X1, X2, …, X10 are control variables; $\varepsilon_{it}$ is a vector of normally distributed errors.

# 5. Results

We present our results in the following order: (1) impulse response functions, (2) Granger causality analysis, (3) reverse causality analysis, (4) boundary conditions analyses, and (5) robustness checks.

## 5.1. Impulse response functions

The impact of a shock[11] to brand equity on vertical integration over time is reflected in the generalized impulse response functions (GIRFs) presented in Fig. 1. The impulse response functions displayed are based on generalized shocks. However, for robustness purposes, we also present the impulse response functions that are based on orthogonalized shocks (see Web Appendix C) obtained from a causal ordering procedure using Cholesky's decomposition of the residuals' matrix (Hamilton, 1994). In addition to the GIRFs, we report the accumulated GIRFs, which represent the cumulative sum of the impact of a shock to one of the proxies of brand equity on the degree of vertical integration in the franchise system over time (see Web Appendix D). As evident in all impulse response function graphs (Fig. 1 and Web Appendix D), a shock to one of the proxies of brand equity (Ad or Media) negatively impacts the degree of downstream vertical integration in the franchise system leading to less vertical integration. Furthermore, the impulse response functions reveal that a shock to one of the proxies of brand equity takes a year or two to start materially impacting the degree of vertical integration in the franchise system. Then, from year 3 onwards, that effect keeps building momentum over time rather than fading away. The Autoregressive Distributed Lag (ARDL) results in Table 4 further confirm this dynamic and provide insights into effect sizes and significance levels. These results provide strong empirical support for hypothesis $H_{1(alt)}$ over its rival $H_1$.

---

[9] When used with individual time series, Unrestricted VARs can produce consistent estimates only when there is a large number of observations. However, firm-level marketing panel data typically have limited time series for each cross-sectional unit (Chakravarty & Grewal, 2011). BVARs overcome the problem of short time-series panels by pooling the data from all cross-sectional units and allowing for heterogeneity in the associations between variables (Holtz-Eakin et al., 1988).

[10] Unrestricted VARs have a problem with overparameterization which limits the number of variables they can handle and leads to inconsistent parameter estimations (Leeper et al., 1996; Maddala, 1992). BVARs overcome this limitation by shrinking the parameter space through the imposition of prior distributions on the parameters (Doan et al., 1984; Leeper et al., 1996) which results in more accurate forecasts (Canova & Ciccarelli, 2004).

[11] A positive shock of magnitude equal to one standard deviation of the residuals of the variable being shocked (the impulse variable).

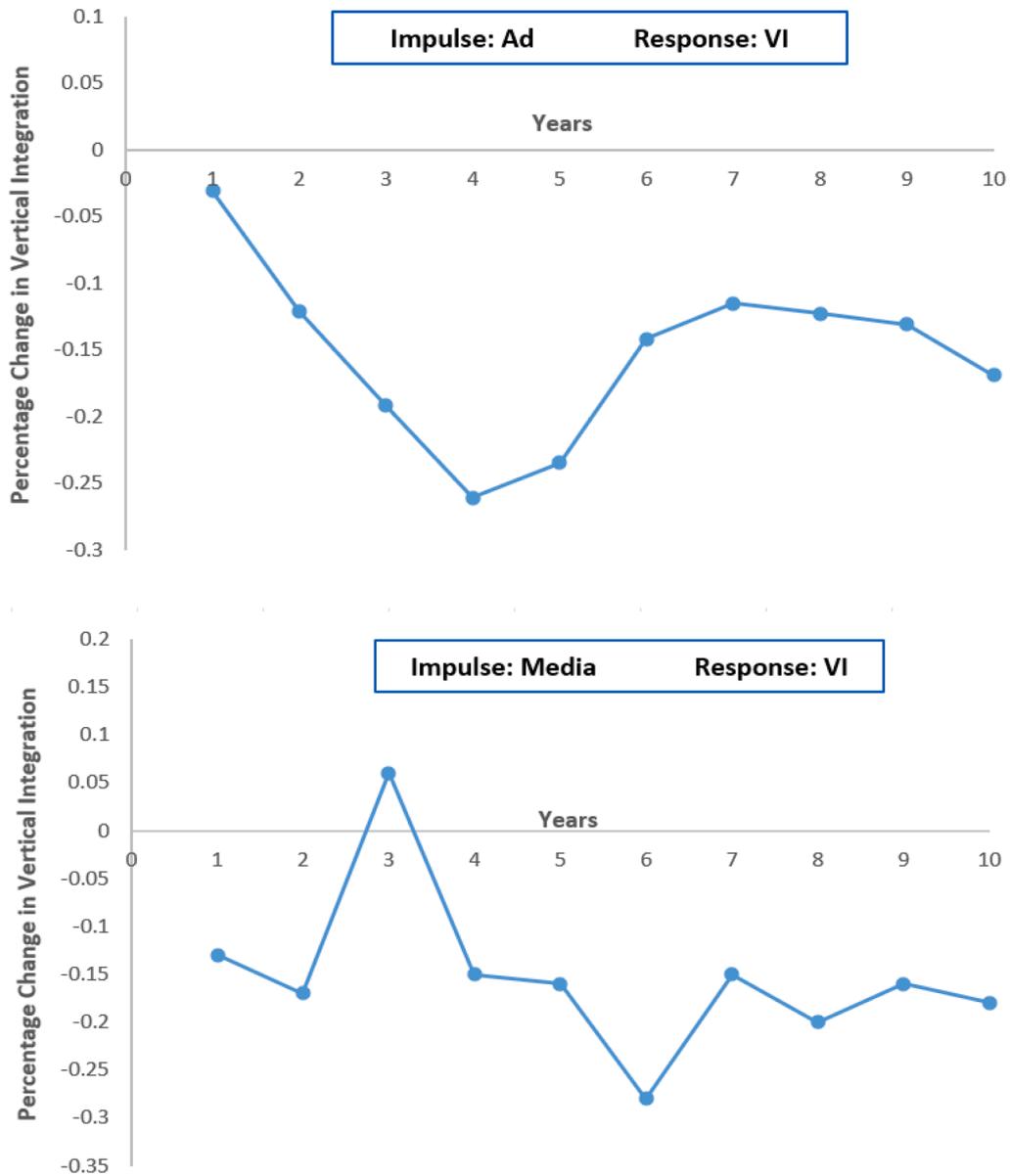

**Fig. 1.** The dynamic impact of brand equity on vertical integration.
Generalized impulse response functions (GIRFs) for the impact of a shock in brand equity on vertical integration. "VI" indicates the degree of vertical integration in the franchise system as reflected by the percentage of company-owned outlets. "Ad" indicates the contractual advertising fee in the franchise system (a proxy of brand equity). "Media" indicates media recognition (a second proxy of brand equity) as reflected by Entrepreneur Magazine's annual ranking of franchise systems. "Impulse: Ad" represents a sudden shock or change in Ad. "Impulse: Media" represents a sudden shock or change in Media. "Response: VI" represents the change in VI in response to those shocks. The graph tracks how VI responds to shocks in Ad and Media.

The GIRFs indicate that the majority of the effect of brand equity on vertical integration tends to be lagging in nature. This is not surprising when we consider the strategic nature of both channel and brand decisions, and the fact that governance adjustment is a time-consuming process that demands significant resource allocation and careful execution. Previous research has documented similar trends (e.g., Mela et al., 1997) and established that, in general, only a small portion of the total impact of brand equity materializes in the short run, while the bulk of the impact is often realized in the future (Aaker & Jacobson, 1994; Mizik, 2014).

### 5.2. Granger causality

A distinctive feature of the VAR models is their ability to assess a certain form of causality known as Granger causality (Granger,

**Table 4**
Results of the autoregressive distributed lag (ARDL) models.

| | Panel A Dependent Variable: Vertical Integration | | Panel B Dependent Variable: Advertising Fee | | Panel C Dependent Variable: Media Recognition | |
| --- | --- | --- | --- | --- | --- | --- |
| | β | SE | β | SE | β | SE |
| **Degree of Vertical Integration Lags** | | | | | | |
| $VI_{t-1}$ | 0.0004 | (0.0022) | 0.0038 | (0.0024) | 0.0704 | (0.1466) |
| $VI_{t-2}$ | 0.0019 | (0.0022) | 0.0028 | (0.0024) | 0.0436 | (0.1452) |
| $VI_{t-3}$ | 0.0003 | (0.0022) | 0.0015 | (0.0024) | 0.1054 | (0.1454) |
| $VI_{t-4}$ | −0.0001 | (0.0022) | 0.0054 | (0.0024)** | −0.0827 | (0.1448) |
| $VI_{t-5}$ | −0.0046 | (0.0022)** | −0.0049 | (0.0024)** | −0.0828 | (0.1458) |
| **Advertising Fee Lags** | | | | | | |
| $\Delta Ad_{t-1}$ | −0.0003 | (0.0183) | 0.0192 | (0.0200) | −1.2010 | (1.2115) |
| $\Delta Ad_{t-2}$ | −0.0217 | (0.0224) | −0.0026 | (0.0245) | −2.1316 | (1.4875) |
| $\Delta Ad_{t-3}$ | −0.0391 | (0.0233)* | −0.0203 | (0.0255) | −3.0952 | (1.5485)† |
| $\Delta Ad_{t-4}$ | −0.0253 | (0.0222) | −0.0357 | (0.0242) | −2.9345 | (1.4755)† |
| $\Delta Ad_{t-5}$ | 0.0100 | (0.0180) | −0.0019 | (0.0197) | −1.6470 | (1.1969) |
| **Media Recognition Lags** | | | | | | |
| $\Delta Media_{t-1}$ | −0.0632 | (0.0933) | −0.0707 | (0.1021) | 37.6557 | (6.1948)† |
| $\Delta Media_{t-2}$ | −0.2353 | (0.0990)** | −0.1695 | (0.1081) | 61.0653 | (6.5813)† |
| $\Delta Media_{t-3}$ | −0.2855 | (0.1004)*** | −0.2811 | (0.1100)** | 56.4789 | (6.6635)† |
| $\Delta Media_{t-4}$ | −0.3210 | (0.0986)*** | −0.1002 | (0.1079) | 58.6124 | (6.5500)† |
| $\Delta Media_{t-5}$ | −0.1005 | (0.0934) | −0.0936 | (0.1022) | 46.7255 | (6.1998)† |
| **Control Variables** | | | | | | |
| Business Development Time | 0.0098 | (0.0032)*** | −0.0132 | (0.0035)† | −0.5186 | (0.2142)** |
| Chain Age | −0.0004 | (0.0034) | 0.0080 | (0.0037)** | −1.4841 | (0.2255)† |
| Geographic Scope | 0.0574 | (0.0733) | −0.0386 | (0.0802) | −17.6253 | (4.8643)† |
| Financing Support | −0.2026 | (0.0804)** | 0.1498 | (0.0883)* | 3.7771 | (5.3353) |
| Chain Size | −0.1906 | (0.0328)† | 0.2132 | (0.0358)† | 77.7877 | (2.1841)† |
| Industry Effect | −0.0075 | (0.0030)** | −0.0098 | (0.0033)*** | −0.1509 | (0.2019) |
| Incentives | 0.0717 | (0.0805) | −0.0051 | (0.0881) | 9.0349 | (5.3453)* |
| Royalty Rate | 0.0270 | (0.0105)** | 0.0435 | (0.0114)† | −1.2275 | (0.6939)* |
| Selection | 0.0814 | (0.0618) | 0.4911 | (0.0676)† | −5.0301 | (4.1030) |
| Socialization | 0.1139 | (0.0229)† | 0.1391 | (0.0251)† | 1.1618 | (1.5227) |
| Intercept | 0.6317* | (0.3379)† | −2.6545 | (0.3705)† | −47.4539 | (22.4621)** |
| Adjusted R-square | 9.26 % | | 8.61 % | | 50.5 % | |
| Included obs. | 1822 | | 1792 | | 1822 | |
| Akaike Criterion | 4224.3 | | 1556.5 | | 1779.4 | |
| Schwarz Criterion | 4389.3 | | 1712.8 | | 1929.3 | |

Robust standard errors are in parentheses.

\* $p < 0.1$;
\*\* $p < 0.05$;
\*\*\* $p < 0.01$;
† $p < 0.001$.

1969). Granger causality is a form of predictive causality that relies on a set of Wald tests to investigate whether (a) the cause is correlated with the effect; (b) the cause precedes the effect, and (c) the cause carries a significant predictive ability about the future values of the effect (Granger, 1980). Following previous research (Borah & Tellis, 2016; Kang et al., 2016; Zellner, 1962), we conduct a set of Wald tests on the estimates of the Autoregressive Distributed Lag (ARDL) model estimated earlier to test for Granger causality (see Table 4, Panel A). The results reveal that both proxies of brand equity - Advertising Fee ($F = 2.34$; $p < 0.1$; $n = 1822$) and Media Recognition ($F = 4.78$; $p < 0.01$; $n = 1822$) - Granger-cause vertical integration.

### 5.3. Reverse causality analysis

We explore the possibility of reverse causality in the relationship under examination. Having established that brand equity Granger-causes vertical integration, we investigate Granger causality in the opposite direction i.e., whether vertical integration has a causal influence on brand equity. We find evidence of reverse causality for one of the proxies of brand equity (Adverting Fee). To get a better understanding of the dynamics of the effect in each direction, we turned to the GIRF graphs. The GIRFs clearly demonstrate that the impact of brand equity on vertical integration (Fig. 1) is more pronounced, persistent, and powerful than that of vertical integration on brand equity (Fig. 2). The ARDL results for the reverse causality analyses are available in Table 4 (see panels B & C).

### 5.4. Boundary conditions analyses

To examine the moderating influences of the factors discussed earlier on the causal link between brand equity and vertical integration in franchise systems, we adopt the standard approach of splitting the sample into subsamples and running the BPVARX model on these subsamples (Colicev et al., 2019). We split the sample based on the five variables discussed earlier which are chain age, chain size, financing support, geographic scope, and industry.[12] First, contrary to the theorized effect ($H_2$), firm resources positively moderate the negative link between brand equity and vertical integration (the effect is stronger in franchise systems with more resources), which refutes the resource-based argument discussed earlier – i.e., the availability of resources for one activity (building brand equity) reduces that for the other (vertically integrating the channel).[13] As shown in Fig. 3, firm resources (operationalized as firm age) amplify the negative impact of brand equity on vertical integration for both proxies of brand equity. The same applies for the other proxy of firm resources, financing support.[14] Second, regarding industry effects, we follow the approach described in Perrigot (2006) and split the sample into retail-focused and service-focused franchises.[15] The results show that the effect is weaker in retail franchises[16] than in services franchises in support of H5. Third, as theorized, we observe that the effect is weaker in international franchises than in domestic franchises in support of H3. Finally, we did not find any moderating effect for chain size (H4) on the link between brand equity and vertical integration in franchise systems.[17] The GIRFs and accumulated IRFs for all subsamples are available in Web Appendix E.

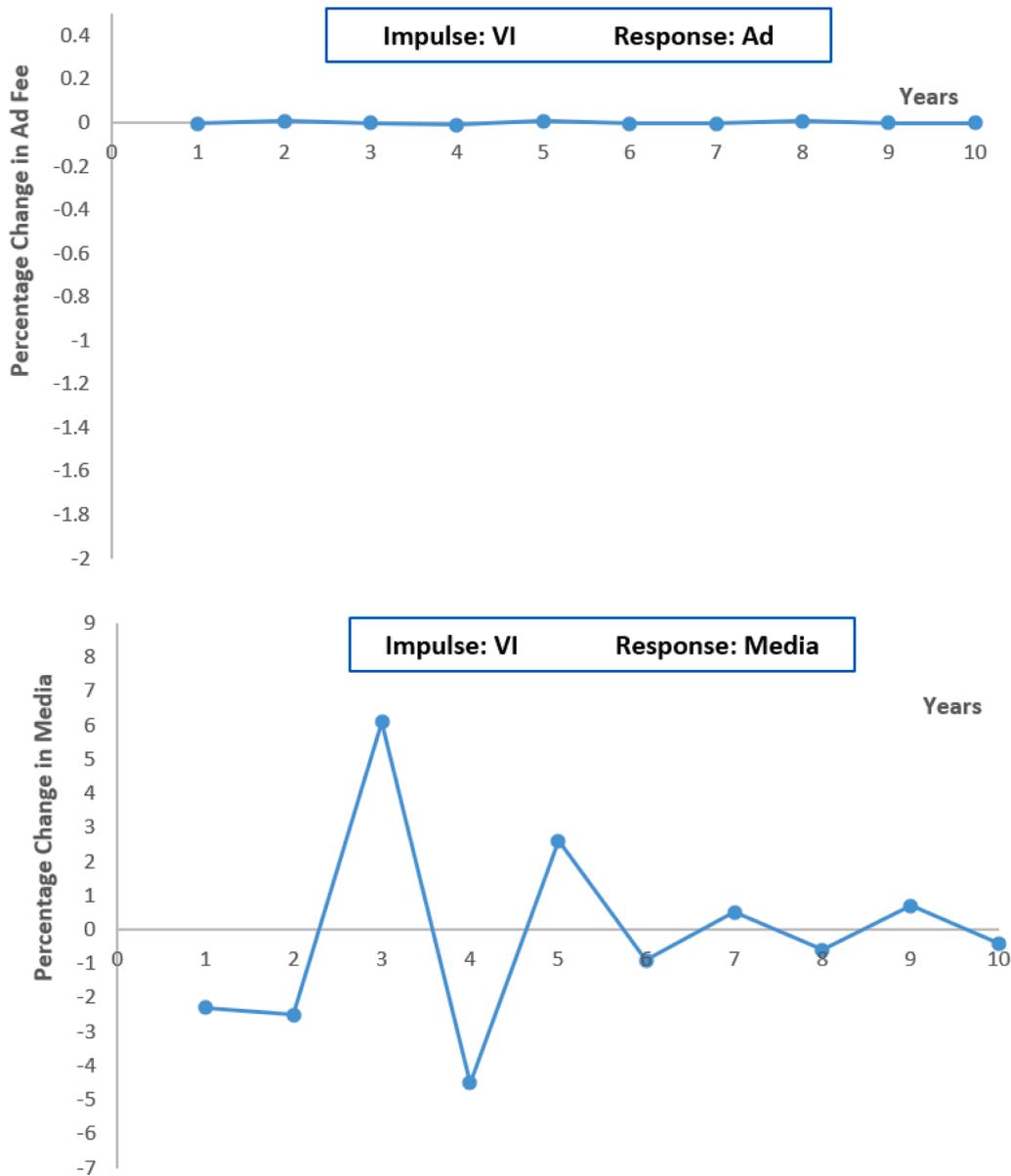

**Fig. 2.** The dynamic impact of vertical integration on brand equity.
Generalized impulse response functions for the effect of a shock in vertical integration on brand equity. "VI" indicates the degree of vertical integration in the franchise system as reflected by the percentage of company-owned outlets. "Ad" indicates the contractual advertising fee in the franchise system (a proxy of brand equity). "Media" indicates media recognition (a second proxy of brand equity) as reflected by Entrepreneur Magazine's annual ranking of franchise systems. "Impulse: VI" represents a sudden shock or change in VI. "Response: Media" represents the change in Media in response to that shock. "Response: Ad" represents the change in Ad in response to that shock. The graph tracks how brand equity (measured by Ad and Media) responds to shocks in VI.

---

[12] Whenever we ran the BPVARX model on the subsamples resulting from sample splitting around one variable, we maintained the other four boundary conditions' variables as control variables in the model.

### 5.5. Robustness checks

To further validate our results and exclude some potential alternative explanations and statistical biases, we conducted several robustness checks. We discuss them below.

*Alternative proxy of brand equity.* To test the robustness of our results, we replaced Media Recognition with an alternative proxy that also serves as an indication of brand equity, Franchise Fee or the Upfront Franchise Fee,[18] which is the initial one-time payment paid by the franchisee to the franchisor to use its brand and business format (Kaufmann & Dant, 2001; Lanchimba et al., 2018). The results remained consistent as reflected in the GIRFs in Web Appendix F and the accumulated GIRFs in Web Appendix G.[19]

*Alternative PVARX specifications.* To assess the robustness of our model, we ran a two-regime switching PVARX model (Chevallier, 2011). Switching VAR models relax the classic VAR assumptions of linearity and constant parameters and structure over time, which allows them to capture nonlinear relationships and changes in the relationship between endogenous variables during different time periods. Our results remained consistent (see the GIRFs in Web Appendix H and the accumulated GIRFs in Web Appendix I).

We also ran an unrestricted PVARX model. As discussed earlier, unrestricted VARs are prone to the problem of over-parameterization (Maddala, 1992, p.602) which leads to forecasts with large standard errors and imprecise coefficient estimates (Canova, 2007, p.373). In addition to that, they do not work well with short panels, which is often the case in most marketing strategy panel data where we often have many individual units and few time periods (Chakravarty & Grewal, 2011). This makes them less consistent than their Bayesian counterparts, which have consistently produced better results and maintain a strong forecasting record (Maddala, 1992). That said, we ran an unrestricted PVARX model to see whether our results remain consistent and to further confirm the temporal causation argument. Our results remained consistent (see the GIRFs in Web Appendix J and the accumulated GIRFs in Web Appendix K).

*Alternative prior distribution.* In this study, we followed previous research (e.g., Chakravarty & Grewal, 2011) in our choice of the Bayesian prior distribution and applied a Wishart prior. To test the robustness of our results against alternative prior distributions, we ran the model using a Minnesota prior. The results remained consistent (see Web Appendix L).

*Orthogonalized vs. generalized IRFs.* In our results section, we followed extant research in marketing (e.g., Borah & Tellis, 2016; Kang et al., 2016) and presented the IRFs that are based on generalized shocks. To further validate our results, we produced the IRFs that are based on orthogonalized shocks obtained from a causal ordering procedure using Cholesky's decomposition of the residuals matrix. We found them to be very similar to the generalized IRFs. We report them in Web Appendix D.

*Alternative lag length selection criteria.* The choice of the optimal lag length for our model was based on the results provided by three-out-of-the-five lag length selection criteria we used (Akaike Information Criterion, Akaike's Final Prediction Error Criterion, and Sequential Modified LR Test Statistic). To test the robustness of our results to alternative lag length selection criteria, we ran the BPVARX model on three lags (L = 3) as suggested by the remaining two selection criteria (Schwarz Information Criterion, Hannan-Quinn Information criterion). The results remained consistent (see Web Appendices M and N).

---

[13] This result offers a refutation of the ownership redirection hypothesis, especially in franchise systems with robust brand equity, which are typically the ones where ownership redirection is expected to occur. The ownership redirection hypothesis posits that franchisors use franchisees to enter markets, expand the chain, and cultivate consumer acceptance of the brand, only to later appropriate brand equity by terminating or otherwise revoking the franchisees' franchising rights once the franchisor possesses sufficient resources. Our results indicate that franchisors with strong brand equity tend to increasingly rely on franchised outlets as their financial resources grow.

[14] Due to space limitations, the remaining IRF graphs pertaining to the boundary conditions analyses are reported in Web Appendix E.

[15] Two coders independently mapped the 44 industries in the Bond's data to their respective classifications. Intercoder reliability was high (above 89%).

[16] For further validation, we split the sample using two other approaches: a three-category classification into accommodation, food, and others (Hsu et al., 2017) and an 8-industry categorization (Gillis et al., 2020). These two additional approaches did not yield any industry-level effects.

[17] A limitation of split-sample testing of moderation effects is that it doesn't account for the possibility of the main effect of the moderator.

[18] The natural logarithm of the upfront franchise fee in USD.

[19] Evidence of Granger causality was not detected for this proxy (Franchise Fee).

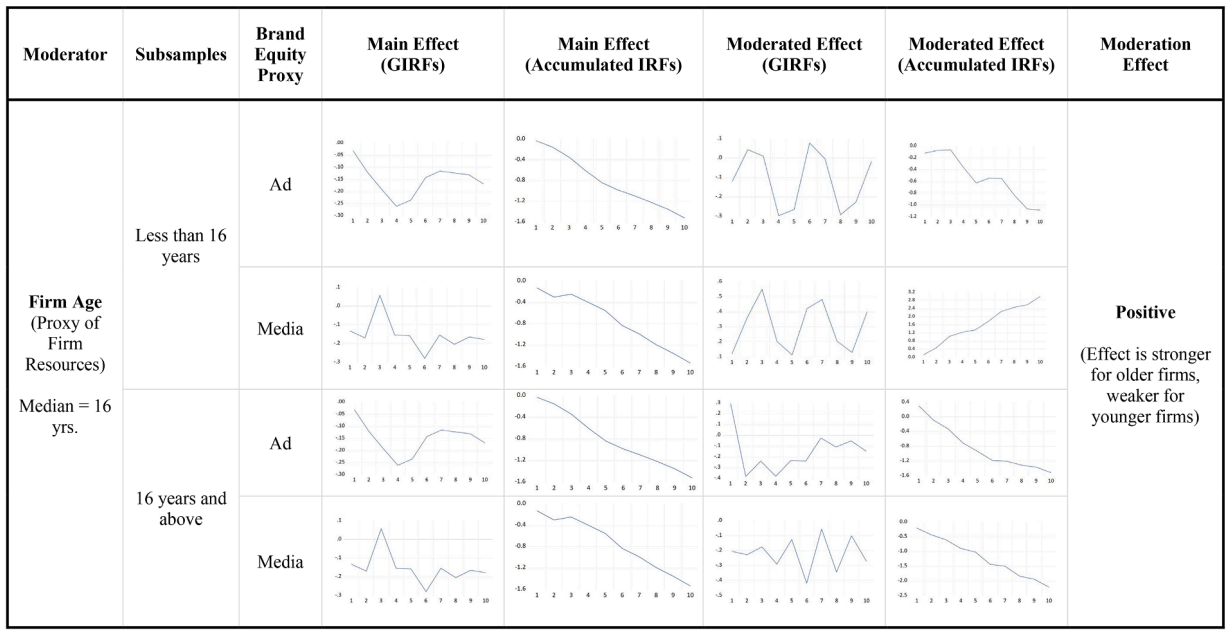

| Moderator | Subsamples | Brand Equity Proxy | Main Effect (GIRFs) | Main Effect (Accumulated IRFs) | Moderated Effect (GIRFs) | Moderated Effect (Accumulated IRFs) | Moderation Effect |
|---|---|---|---|---|---|---|---|
| **Firm Age** (Proxy of Firm Resources) Median = 16 yrs. | Less than 16 years | Ad | | | | | **Positive** (Effect is stronger for older firms, weaker for younger firms) |
| | | Media | | | | | |
| | 16 years and above | Ad | | | | | |
| | | Media | | | | | |

**Fig. 3.** Impulse response functions (IRFs) for the moderation effect of firm resources on the link between brand equity and vertical integration. "Ad" indicates the contractual advertising fee in the franchise system (a proxy of brand equity). "Media" indicates media recognition (a second proxy of brand equity) as reflected by Entrepreneur Magazine's annual ranking of franchise systems. The impulse variables are Ad and Media. The response variable is VI, the degree of vertical integration in the franchise system as reflected by the percentage of company-owned outlets. The graph tracks how VI responds to shocks in Ad and Media, demonstrating the positive moderation effect of firm resources (proxy: firm age) on the link between brand equity and vertical integration.

*BPVAR vs. BPVARX.* To test the robustness of our results against the potential influences of some control variables, we ran a BPVAR model i.e., the model with endogenous variables only, excluding all controls. The results remained consistent for both prior distributions.[20] We also tested for time-dependent unobserved heterogeneity and found no evidence of it in our system.[21]

*Selection bias.* To test the robustness of our results to selection bias, we extracted the balanced sub-panel from our unbalanced panel and ran the same model on it (Balestra and Nerlove, 1966). The results remained consistent for both lag lengths (L = 5 and L = 3) and under both prior distributions, the Wishart and Minnesota.

*Sample excluding observations with 100 % and 0 % vertical integration.* To further test the robustness of our results, we ran the model on three subsamples: the sample excluding observations with 0 % vertical integration, the sample excluding observations with 100 % vertical integration, and the sample excluding observations with either 0 % or 100 % vertical integration. The results remained consistent for all three models and under both prior distributions, Wishart and Minnesota.

*Outliers.* To test the robustness of our results to the exclusion of extreme values, and to further confirm the presence of the effect, we ran the BPVARX model on four trimmed subsamples (excluding the ± 5 percentiles and the ±10 percentiles). The results remained consistent under both prior distributions, Wishart and Minnesota.

## 6. Discussion

In our endeavor to calibrate the impact of brand equity on vertical integration in franchise systems, we (a) reviewed existing research and presented a hypothesis that represents the majority of available evidence on the topic, (b) drew on power-dependence theory (Emerson, 1962) and the theory of self-enforcing contracts (Klein, 1985; 2002) and presented a rival hypothesis, (c) tested these two competing hypotheses using a Bayesian Panel Vector Autoregressive (BPVARX) model on a large panel data set, and (d) conducted a boundary conditions analysis. Our results show a powerful but lagging inverse effect for brand equity on vertical integration such that higher brand equity leads to less vertical integration in a franchise system. This negative effect is weaker in franchise systems with international presence and in retail-focused franchises, and stronger in franchise systems with more financial resources.

---

[20] Impulse Response Functions for this model and all remaining models are available on request.

[21] We tested for time-dependent unobserved heterogeneity using five approaches: linear time trends (using Year as a continuous variable), event-based year-grouping dummies (during vs. before great recession), economic performance-based year-grouping dummies (during low vs. non-low economic growth period), and their combinations. The results remained unchanged across all five models. Including year dummies in a high-lag PVAR system on a short or unbalanced panel often results in overparameterization and multicollinearity, rendering a PVAR system unstable/unspecified (Baltagi, 2005; Hamilton, 1994), as demonstrated in our analysis.

## 6.1. Theoretical contributions

Our study makes several contributions to marketing theory. First, we advance the franchising and retailing literatures by providing novel insights into the decades-old debate on the impact of brand equity on vertical integration in franchise systems. In doing so, we depart from previous research in several ways. Theoretically, while the majority of existing research has examined the topic through the lenses of transaction cost theory and agency theory, we ground our investigation in the power-dependence theory and the theory of self-enforcing contracts. Methodologically, this is the first study to explore causal, lagged, reverse, and moderation effects. Moreover, we not only demonstrate the negative causal effect of brand equity on vertical integration and identify key moderating factors, but more importantly, we uncover the intricate dynamics of this effect. In particular, we reveal the temporal progression of the effect, which begins slowly but continues building momentum, with only a small portion of the effect materializing in the short run and the majority of the effect manifesting in the long run. Moreover, we find evidence of a less pronounced but present reciprocal effect. These insights were facilitated by the use of a Bayesian Panel Vector Autoregressive model, which allowed us to investigate lagged effects while controlling for endogeneity and firm-level heterogeneity, examine temporal causation, and assess reverse causality. Our study, one of the few in the retailing literature to utilize this approach, presents an example of how dynamic research methods can address the limitations of prior studies to provide more nuanced, comprehensive, and in-depth insights into a number of marketing phenomena (see Tables 1 and 2).

Second, our findings contribute to the broader channels' literature. The influence of brand equity on distribution channels, as a general phenomenon, is still a relatively under-researched topic in marketing despite practitioners' and scholars' recognition of the crucial role brands play in distribution channels (Hoeffler & Keller, 2003). Indeed, extant research in marketing on the impact of brand equity on distribution focuses mostly on channel coordination and primarily investigates how an upstream firm's brand equity influences the tactical behavior of its downstream channel partners. Research on the impact of brand equity on channel governance/structure is relatively limited. Our extensive review of the channels' literature has identified two theoretical arguments about the impact of brand equity on channel governance but no empirical evidence. Both arguments are in line with our view on the topic. The first is by Ghosh and John (1999) who posit that when brand equity is high, the firm is more capable of using market governance, whereas weaker brands "handicap" the firm from doing so. The second argument is by Coughlan et al. (2006, p.351) who argue that when brand equity is high, vertical integration into distribution is not only unnecessary but rather "wasteful." Aside from these two theoretical arguments, marketing theory seems relatively silent on the subject despite its recognition that "brand equity influences governance directly." (Ghosh & John, 1999, p.140). Hence, our findings, albeit from the franchising context, provide much needed evidence that supports these two theoretical views on the impact of brand equity on vertical integration. In doing so, we scratch the surface of an important relationship that is recognized by both practitioners and scholars, yet not sufficiently researched in marketing. Table 5 provides a summary of existing research in marketing on the impact of brand equity on distribution and illustrates how our study advances that body of research.

**Table 5**
Overview of research in marketing on the impact of brand equity on distribution channels.

| Study | Context | Channel Coordination/ Governance | Channel Member Under Study | Brand Equity Operationalization | Key Relevant Findings |
|---|---|---|---|---|---|
| Montgomery (1975) | Grocery Store / Supermarket | Channel Coordination | Brand Seller (Downstream firm) | Advertising | Stronger brands have better chance of being accepted at stores and in gaining shelf-space. |
| Farris et al. (1989) | Grocery Store / Supermarket | Channel Coordination | Brand Seller (Downstream firm) | % of survey subjects who would choose the brand over rivals | Retailers provide better in-store merchandising and stocking to stronger brands. |
| Fader and Schmittlein (1993) | Grocery Store / Supermarket | Channel Coordination | Brand Seller (Downstream firm) | Market share | Stronger brands have higher availability at retailers. Retailers who carry few brands carry those with higher brand equity. |
| Lal and Narasimhan (1996) | Not Applicable (Analytical model) | Channel Coordination | Brand Seller (Downstream firm) | Advertising | Retailers are willing to accept lower margins on stronger brands because they see them as drivers of store traffic. Retailers are more likely to advertise stronger brands because customers use them to gauge the store's overall price levels. |
| Bell et al. (1999) | Grocery Store / Supermarket | Channel Coordination | Brand Seller (Downstream firm) | Average number of purchases of the brand per consumer | During promotions, retailers stockpile stronger brands more than weaker brands. |
| Besanko et al. (2005) | Grocery Store / Supermarket | Channel Coordination | Brand Seller (Downstream firm) | Market share | Stronger brands receive higher promotion pass-through (by retailers) than weaker brands. |
| This Study | Franchising, 44 industries | Channel Governance | Brand Owner (Upstream firm) | Advertising, Brand Ranking, Franchise Fee | As brand equity increases, franchisors rely less on downstream vertical integration in governing their franchise chains. This is because brand equity functions as safeguarding asset that solves many channel governance problems through contractual self-enforcement, which provides an effective safeguard against franchisees' opportunism and alleviates the need for vertical bureaucratic control through vertical integration. |

Particularly, our findings add to the rich channel governance literature by proposing brand equity as an alternative governance mechanism that enables the firm to govern its distribution channel effectively without the need for vertical integration. The channels governance literature provides several examples of such alternative governance mechanisms that address governance issues without the need for vertical integration. For instance, Anderson and Weitz (1992) discuss the role of pledges in building trust and commitment between partners, effectively reducing opportunistic behavior. Stump and Heide (1996) emphasize the importance of careful partner selection, tailored incentive design, and rigorous monitoring to ensure alignment of interests and performance standards. Heide and John (1992) highlight the efficacy of relational norms, such as flexibility, information sharing, and long-term orientation, in fostering cooperative behavior and mitigating conflicts within channel relationships.

Third, our findings extend the brand equity literature by identifying a new strategic role for brand equity that goes beyond customers, competitors, employees, and shareholders to reach channel partners. Brand equity is a vital strategic asset that can help a franchisor in governing its franchise system by acting as a "safeguarding asset" that bridles franchisees' opportunism through contractual self-enforcement, and thus alleviate the need for deep vertical integration.

Fourth, we contribute to an under-researched stream in marketing strategy that is concerned with understanding the interactions among marketing mix elements. Most marketing strategy research focuses on a particular element of the marketing mix such as channel, brand, or pricing (Srinivasan, 2006). But, in reality, firms craft their marketing strategy as an intertwined whole and consider synergies, tradeoffs, and interdependencies among marketing mix elements (Gatignon & Hanssens, 1987). Hence, exploring such interactions and interdependencies is crucial for both scholarship and practice. In this spirit, we add to the marketing interactions research stream (e.g., Gatignon & Hanssens, 1987; Srinivasan, 2006) by examining the dynamic association between brand and distribution.

### 6.2. Managerial implications

Our study offers a number of valuable, actionable insights that can help senior managers in their strategic decision making. First, for channel managers of franchise systems, our research provides novel insights into the long-standing question of whether to vertically integrate or not to integrate. Despite the resurging interest in vertical integration amongst franchising and retailing practitioners (and in line with a significant body of scholarly work in marketing), we advise against unnecessary vertical integration especially in situations where the franchise enjoys a moderate to high level of brand equity. Whereas the temptation of control might compel some managers to pursue vertical integration, it is a costly, risky investment that demands large resource commitments that often outweigh the foreseeable gains of such a venture. Therefore, it should only be considered in situations of low or unstable brand equity and only after exhaustive scrutiny. As evidenced in this study, as the brand equity of a franchise system increases, the franchisor leans more on its brand to safeguard itself against franchisees' opportunism and to govern its franchise system effectively without the need for deep vertical integration.

Second, we offer brand managers of franchise systems useful empirical evidence that can help them in selling brand building initiatives to senior management and other stakeholders. Brand managers are often faced with the challenging task of selling brand building initiatives to senior management and other organizational stakeholders due to the intangibility and long-term nature of the beneficial outcomes of such initiatives. This study can help brand managers of franchise systems in gaining organizational support for brand building activities by shedding light on an under-recognized, strategic benefit for brand equity – by strengthening its brand equity, the franchisor mitigates many channel governance issues through contractual self-enforcement which reduces the need for deep vertical integration. This can make the challenging task of selling brand building initiatives to various organizational stakeholders much easier. Additionally, the documented lagged nature of the impact of brand equity on vertical integration underscores the need for firms to adopt a long-term perspective when evaluating the returns on brand investments as these returns often accrue gradually and compound over time. As evidenced in this study, brand equity can have profound and positive influences on certain long-term marketing decisions (e.g., mitigating the need for the costly and risky vertical integration of the channel). However, such positive influences become apparent only over time which emphasizes the importance of a strategic focus when it comes to investments in brand equity.

Third, senior marketing executives (e.g., CMO) of franchise systems, often find themselves dealing with a capital allocation conundrum where different marketing strategies (e.g., invest in direct distribution, invest in strengthening the brand) compete for constrained financial resources. Remarkably, whereas "gaining organization support and resources for brand building activities is often difficult, even with a consensus that brands are strategically important to the organization," (Biel & Aaker, 1993, p.333) the desire for control compels many executives to pursue vertical integration, even though it is a costly, risky investment that demands large resource commitments that often outweigh the foreseeable gains of such a venture (Bateman, 2016). Our results suggest that, in franchise systems, investments in brand equity may offer a lower risk/reward ratio and a better hedge against uncertainty than many other investments. This is because brand investments can be seen as "dual investments" directly in the brand and indirectly in the channel (brand investments significantly mitigate the channel governance problem which reduces the need for deep involvement in direct distribution).

### 6.3. Limitations and future research directions

Our work has some limitations that offer useful avenues for future research on the topic. First, despite the broadness of our context (44 industries) and its adequacy to the research question, and despite our reliance on well-established measures from the literature, the generalizability of our results could be further improved through the convergence of findings from various studies using alternative measures in different contexts. Hence, future research could examine the robustness of our results to different contexts and measures. This will provide an even more nuanced understanding of the relationship between brand equity and vertical integration in franchising while assessing the consistency of our findings across diverse contexts. More importantly, future research could examine whether our findings can be generalized to non-franchising retail settings. Second, due to data limitations and because the focus of this study is to establish the impact of brand equity on vertical integration, we do not test the underlying process by which higher brand equity leads to less downstream vertical integration in franchise systems. Moreover, doing so would require measuring such concepts as dependence, self-enforceability, incentives, safeguards, replaceability, termination sanction, and opportunistic behavior. Such constructs naturally lend themselves to survey data and cannot be easily captured by the data sources we use in this study. Hence, future research may apply different measurements, use alternative contexts, and test some of the links in the underlying process. In addition to that, future research could examine whether the same effect holds for backward vertical integration, particularly because there is some anecdotal evidence suggesting that retailers such as Walmart and Amazon rely on their brand equity in governing their supply chain without the need to acquire their suppliers. Finally, as we highlighted earlier, the effect of brand equity on various aspects of distribution strategy is a much under-researched topic in marketing. Hence, future research is urged to further explore this research area and examine other facets of the influence of brand equity on distribution strategy.

### Declaration of generative AI and AI-assisted technologies in the writing process

During the preparation of this article, some of the authors used ChatGPT in order to conduct grammar checking, paraphrase sentences, verify citation style, search for published papers, summarizing text, analyzing text, and other writing- or style-related assistance. After using this tool, the authors carefully reviewed and edited the content as needed and take full responsibility for the content of the publication.

### Acknowledgments

The authors are thankful for the valuable feedback provided by Rob Palmatier, Venky Shankar, and participants at the Theory + Practice in Marketing (TPM) Conference, the AMA Winter Academic Conference, the European Marketing Academy (EMAC) Conference, the ISBM Academic Conference, the International Society of Franchising (ISoF) Conference, and other conferences and seminars. Funding for this research was provided by the Institute for the Study of Business Markets (ISBM) at Penn State. However, the contents of this research reflect the views of the researchers who are solely responsible for the facts and the accuracy of the data presented herein. This research is supported by funding for Manish Kacker from the Social Sciences and Humanities Research Council of Canada (SSHRC). Mohammad Kayed is grateful for the generous financial support from the Social Sciences and Humanities Research Council (SSHRC) of Canada, award number 767-2016-2275.

### Supplementary materials

Supplementary material associated with this article can be found, in the online version, at doi:10.1016/j.jretai.2025.01.007.